\documentclass[aps,prl,twocolumn,superscriptaddress,nofootinbib,preprintnumbers,10pt]{revtex4-1}

\usepackage{booktabs}
\usepackage{hyperref}
\usepackage{multirow} 
\usepackage{numprint}
\usepackage{enumitem}
\usepackage{mathrsfs}
\usepackage{graphicx}
\usepackage{multirow}
\usepackage{makecell}
\usepackage{verbatim}
\usepackage{color}
\usepackage{comment}
\usepackage{epstopdf}
\usepackage{float}
\usepackage{amsmath}
\usepackage{amssymb}
\usepackage{tikz}
\usepackage{wrapfig}
\usepackage{tikz-feynman}
\usepackage{verbatim}
\pdfoutput=1
\usepackage[utf8]{inputenc}
\usepackage{graphicx,amsmath,amsfonts,amssymb,aas_macros,slashed,float}
\usepackage{bbold}
\usepackage{numprint}
\usepackage{enumitem}
\usepackage{graphicx,color,amsmath}
\usepackage{hyperref}
\usepackage{hyperref}
\usepackage{url}
\usepackage{slashed,stmaryrd}

\newcommand*{\dif}{\mathop{}\!\mathrm{d}}
\allowdisplaybreaks
\makeatletter
\newcommand{\xRightarrow}[2][]{\ext@arrow 0359\Rightarrowfill@{#1}{#2}}
\makeatother
\usepackage{pgfplots}

\usepackage{bm}
\usepackage{mathtools}

\usetikzlibrary{positioning,arrows,patterns}
\usetikzlibrary{decorations.markings}
\usetikzlibrary{calc}

\begin{document}
	\title{Enhancement of the screening effect in semiconductor detectors in the presence of the neutrino magnetic moment}
	\author{Yu-Feng Li}
	\email{liyufeng@ihep.ac.cn}
	\affiliation{Institute of High Energy Physics, Chinese Academy of Sciences, Beijing 100049, China}
	\affiliation{School of Physical Sciences, University of Chinese Academy of Sciences, Beijing 100049, China}
	\author{Shuo-yu Xia}
	\email{xiashuoyu@ihep.ac.cn}
	\affiliation{Institute of High Energy Physics, Chinese Academy of Sciences, Beijing 100049, China}
	\affiliation{School of Physical Sciences, University of Chinese Academy of Sciences, Beijing 100049, China}
 
	\begin{abstract}
	The theoretical framework of the neutrino electron excitation at low energies including the screening effect in semiconductor detectors is developed for the first time, both in the Standard Model and in the presence of the neutrino magnetic moment. We explore the contribution of the screening effect of semiconductors to the neutrino electron excitation based on the linear response theory and calculate the corresponding numerical results. We show that excitation rates from the neutrino magnetic moment are dramatically enhanced {by} the screening effect and the sensitivity can be significantly improved to the level of $10^{-13}\,\mu_{\rm B}$, much better than the current {laboratory and astrophysical} limits.
	\end{abstract}
	\maketitle                                                                                              
\paragraph{\textbf{Introduction.}}
Neutrino interaction~\cite{Formaggio:2012cpf} is one of the most important ingredients of the neutrino observation, among which the elastic neutrino electron scattering process has been widely used in the detection of solar neutrinos~\cite{Kamiokande-II:1989hkh,Super-Kamiokande:1998qwk,Borexino:2007kvk}, reactor neutrinos~\cite{Amsler:2002tu,TEXONO:2002pra,Beda:2007hf} and accelerator neutrinos~\cite{CHARM:1988tlj,LSND:2001akn}, and played important roles in studying fundamental properties of neutrino oscillations and new physics beyond the Standard Model (SM). {For the state-of-the-art semiconductor dark matter (DM) detectors, such as SENSEI~\cite{SENSEI:2020dpa}, EDELWEISS~\cite{EDELWEISS:2020fxc} and SuperCDMS~\cite{SuperCDMS:2017mbc}}, the {sensitivity has reached sub-keV and} the collective behaviors of electrons can make important contributions to the relative {electron excitations, which are neglected in most of the present applications.}

As a fundamental property of massive neutrinos, the magnetic moment is estimated to be vanishingly small in simple extensions of SM~\cite{PhysRevLett.45.963,Giunti:2014ixa}. However, the neutrino magnetic moment can be significantly enhanced in many models beyond the simplest SM extension~\cite{Bell:2006wi,Bell:2005kz}, and even can reach the testable level of the current limits from laboratory measurements~\cite{Amsler:2002tu,TEXONO:2002pra,Beda:2007hf,Giunti:2014ixa} and astrophysical considerations~\cite{PhysRevLett.111.231301,C_rsico_2014,Carenza:2022ngg}. Note that the recent results from XENONnT have pushed the laboratory limit down to ~$6.4\times10^{-12}\mu_{\rm{B}}$ at the 90\% confidence level~\cite{XENONCollaboration:2022kmb}. Since { the contribution of the neutrino magnetic moment to the neutrino electron excitation rate} is {enhanced by the inverse of the recoil} energy, {semiconductor detectors take significant advantages compared to other popular detectors for a lower energy threshold.}
	
The thresholds of present semiconductor detectors are already capable of going even below 0.1 keV, {where} the conventional scattering theory with non-interacting particle states cannot precisely describe the related physics. The collective behaviors of electrons {at such low energies} in semiconductor detectors, which have attracted various attentions in the fields of DM direct detection~\cite{PhysRevD.102.015017,PhysRevD.101.123012,PhysRevLett.121.101801,PhysRevD.101.076014,GELMINI2020135779,PhysRevD.104.056009}, play an important role in such scenarios. The related underlying physics {including the screening effect} can be well described with the dielectric function of the material and the in-medium effects of the DM-electron excitation are thoroughly investigated in Refs.~\cite{PhysRevLett.127.151802,PhysRevD.104.015031}. 
	
The theory to describe the in-medium screening effect for DM-electron scattering in crystal have already been developed in Ref.~\cite{Liang:2021zkg}.
The energy loss function (ELF), which is defined as the imaginary part of the inverse dielectric function 
{{in the homogeneous electron gas (HEG)}}, is the key to describe the DM-electron excitation. {{To generalize the results of HEG to the isotropic crystal target, an effective form of the inverse dielectric function is taken as 
the average of the diagonal elements of the general matrix of the inverse dielectric function. Moreover, the linear response theory is employed to describe the screened DM-electron scattering as a perturbation exerted onto the electron system.}
 
The purpose of this work is to develop for the first time the general theory of the neutrino-electron excitation in isotropic semiconductors {including the screening effect}, since previous calculations from the conventional scattering theory have underestimated some important features of the electron response under low energies. We generalize {the non-relativistic effective filed theory framework (NR EFT) in Ref.~\cite{Mitridate:2021ctr}} to the neutrino-electron couplings in order to study the effective Lagrangian applied in the non-relativistic neutrino-electron excitation.
In order to illustrate the enhanced response in the contribution of the neutrino magnetic moment, we calculate the constraints from semiconductor detectors using the solar neutrinos, whose sensitivity approaches the level of $10^{-13}\,\mu_{\rm B}$, much better than the current best limit from laboratory and astrophysical probes.

	\paragraph{\textbf{Neutrino Couplings to Non-Relativistic Electrons.}} 
	The elastic neutrino electron scattering (E$\nu$ES) can be described with the following standard Lagrangian 
	\begin{equation}
	\begin{aligned}
	\mathcal{L}_{\rm E\nu ES}=-i\frac{G_F}{\sqrt{2}}\left[\bar{\nu}_{\alpha}\gamma^{\rho}(1-\gamma^5)\nu_{{\alpha}}\right]\left[\bar{\phi}_{e}\gamma_{\rho}(g_{\alpha;\rm{V}}-g_{\alpha;\rm{A}}\gamma^5)\phi_{e}\right]\,,
	\end{aligned}
	\end{equation}  
where $g_{e;\rm{V}}=1/2+2\sin^2\theta_{W}$, $g_{e;\rm{A}}=1/2$ for the electron neutrino and $g_{\mu,\tau;\rm{V}}=-1/2+2\sin^2\theta_{W}$, $g_{\mu,\tau;\rm{A}}=-1/2$ for muon and tau neutrinos.
$L^{\rho}=\bar{\nu}_{\alpha}\gamma^{\rho}(1-\gamma^5)\nu_{{\alpha}}$ denotes the neutrino current and this structure will be kept unchanged in the following calculation until we deal with the square of the scattering amplitude. Since the scattering process we have focused in this work occurs at very low energies, it is reasonable to employ the approximation that the momentum of electrons can be neglected compared to the electron mass and so is the momentum transfer compared to the neutrino energy. Meanwhile, we make the assumption that the semiconductor crystal is isotropic as in most popular detectors with these mediums. Under the above assumption, the neutrino current $L^{\rho}$ can be written as the respective temporal and spatial components
	\begin{equation}
	\begin{aligned}
	&\overline{u}_{\alpha}^{r}(\bm{k}_1)\gamma^{0}(1-\gamma^{5})u_{\alpha}^{r}(\bm{p}_1)
	\approx 2E_{\nu}-2rE_{\nu},\\
	&\overline{u}_{\alpha}^{r}(\bm{k}_1)\gamma^{i}(1-\gamma^{5})u_{\alpha}^{r}(\bm{p}_1)
	\approx -2E_{\nu}\frac{1}{r}.
	\end{aligned}
	\end{equation}
The spatial part of the neutrino current are approximately cancelled due to the spin summation in the amplitude calculation and only the temporal part contributes to the scattering process at the leading order. Then the Lagrangian of E$\nu$ES can be written as the vector and axial-vector parts
	\begin{equation}
	\begin{aligned}
	\mathcal{L}_{{\rm E\nu ES}}
	&=-i\frac{G_F}{\sqrt{2}}g_{\alpha;\rm{V}}L^{0}\bar{\phi}_{e}\gamma_{0}\phi_{e}+i\frac{G_F}{\sqrt{2}}g_{\alpha;\rm{A}}L^{0}\bar{\phi}_{e}\gamma_{0}\gamma^5\phi_{e}+[\bm{L}]\\
	\end{aligned}\,,
	\label{Lagragian_EvES_VA}
	\end{equation}
	where the terms of spatial components are included in $[\bm{L}]$ and will not be considered at the leading order calculation. In the energy range we are concerned in this work, NR EFT is a reasonable approximation~\cite{Penco:2020kvy,Rothstein:2003mp} to describe the neutrino-electron interaction. To match the relativistic theory of E$\nu$ES onto the framework of the NR EFT with effective operators, we employ the techniques from Ref.~\cite{Mitridate:2021ctr} and obtain the leading order non-relativistic { Lagrangian} of E$\nu$ES as

	{	\begin{equation}
		\label{L_EnES_eff}
	\mathcal{L}_{\rm{E}\nu \rm{ES}}=-i\sqrt{2}G_{\rm F}g_{\alpha;\rm{V}}\bar{\nu}_{{\alpha},L}\gamma^{0}\nu_{{\alpha},L}\phi_{+}^{\dagger}\phi_{+}\,,
	\end{equation}
	}where $V_{\rm{E}\nu \rm{ES}}=-i\sqrt{2}G_Fg_{\alpha;\rm{V}}$, {is defined as} the effective non-relativistic potential of E$\nu$ES, {$\phi_{+}$ is the non-relativistic effective operator of the electron field, and its definition can be found in Eq.~(\ref{phiplus}) of Appendix A. Note that this non-relativistic Lagrangian is vector dominant.
 A detailed calculation of the non-relativistic Lagrangian in Eq.~(\ref{L_EnES_eff}) is provided in Appendix A.}
	
	The scattering process between a electron and a neutrino in the presence of the neutrino magnetic moment can be described with the Lagrangian
	\begin{equation}
		\mathcal{L}_{\rm{mag}}=-i\mu_{\nu}\frac{m_{\nu}}{m_e}\frac{4\pi\alpha}{q^2}\left[\bar{\nu}_{{\alpha}}(1-\gamma^5)i\frac{\sigma^{\mu\nu}q_{\nu}}{2m_{\nu}}\nu_{{\alpha}}\right]\left[\bar{\phi}_{e}\gamma_{\mu}\phi_{e}\right]\,,
	\end{equation}
	with $m_{\nu}$ and $m_e$ being masses of the neutrino and electron respectively and $\mu_{\nu}$ being the neutrino magnetic moment. With the Gordon Identity, the above Lagrangian can be separated into two terms
	\begin{equation}
		\begin{aligned}
			\mathcal{L}_{\rm{mag}}=&-i\mu_{\nu}\frac{m_{\nu}}{m_e}\frac{4\pi\alpha}{q^2}\left[\bar{\nu}_{{\alpha}}\gamma^{\mu}(1-\gamma^5) \nu_{{\alpha}}\right]\left[\bar{\phi}_{e}\gamma_{\mu}\phi_{e}\right]\\
			&+i\mu_{\nu}\frac{1}{2m_e}\frac{4\pi\alpha}{q^2}\left[\bar{\nu}_{{\alpha}}(1-\gamma^5){(p_{1}+k_{1})^{\mu}} \nu_{{\alpha}}\right]\left[\bar{\phi}_{e}\gamma_{\mu}\phi_{e}\right]\,,
		\end{aligned}
	\label{NMM_s}
	\end{equation}	
	where $p_1$ and $k_1$ are the four momenta of the initial and final neutrinos respectively and can be approximately considered as the same since the momentum transfer is far less than the neutrino momentum. In this work the neutrino mass can be neglected and the first term will not be considered in the following leading order calculation. The spatial component of the neutrino four momentum will be integrated out as zero due to the isotropic structure of the crystal and we do not need to consider this term as well. As a result, the Lagrangian to describe the contribution of the neutrino magnetic moment at leading order in the isotropic crystal can be written with the temporal component of the second term in Eq.(\ref{NMM_s})
	  \begin{equation}
	  	\begin{aligned}
	  		\mathcal{L}_{\rm{mag}}
	  		&=i\mu_{\nu}\frac{E_{\nu}}{m_e}\frac{8\pi\alpha}{q^2}\left[\bar{\nu}_{{\alpha},L} \nu_{{\alpha},L}\right]\left[\bar{\phi}_{e}\gamma_{0}\phi_e\right]\,.
	  	\end{aligned}
	  \end{equation}
  The structure of the leading order Lagrangian is the same as the vector part of the {standard} E$\nu$ES Lagrangian and it can be written in a similar way { as in Appendix A} with the non-relativistic effective Lagrangian
{  	\begin{equation}
  		\label{L_mag_eff}
  		\begin{aligned}
  			\mathcal{L}_{\rm{mag}}&=i\mu_{\nu}\frac{E_{\nu}}{m_e}\frac{8\pi\alpha}{q^2}\bar{\nu}_{{\alpha},\rm{L}} \nu_{{\alpha},\rm{L}}\phi_{+}^{\dagger}\phi_{+}\\&=V_{\nu,\rm{mag}}(\bm{q})\bar{\nu}_{{\alpha},\rm{L}} \nu_{{\alpha},\rm{L}}\phi_{+}^{\dagger}\phi_{+}\;,
  		\end{aligned}
  	\end{equation}
}where { $V_{\nu,\rm{mag}}(\bm{q})$ is similarly defined as the effective potential} of the contribution of the neutrino magnetic moment.
  
	\paragraph{\textbf{E$\nu$ES in Crystalline Solids.}}
	As mentioned above, to describe the scattering process in the crystal in the context of the linear response theory, it is convenient to regard the effects of the incident particle as a perturbation on the electron system~\cite{Liang:2021zkg}. For the case of electron energy loss spectroscopy (EELS) in the HEG, the effects of incident electron can be described with the following effective Hamiltonian with the Coulomb potential $V_{\rm{cou}}(\bm{Q})$
	\begin{equation}
		\hat{H}_{I}(t)=V_{\rm{cou}}(\bm{Q})\int e^{i \bm{Q}\cdot \bm{x}}\hat{\rho}_{I}(\bm{x},t)e^{i\omega_{p^\prime p}t} \dif^3 x\,,
	\end{equation}
	where $\hat{\rho}_{I}(\bm{x},t)$ is the density operator of the electron and $\omega_{p^\prime p}=E_{e}^\prime-E_{e}$ is the energy difference of the electron. As in Ref.~\cite{Liang:2021zkg}, the averaging and spin summing calculation can be related to a correlation function
	\begin{equation}
		\begin{aligned}
		S^{\hat{H}^{\dagger}_{I}\hat{H}_{I}}(-\omega_{p^\prime p})=	\sum_{i,f}p_{i}\left|<f|\hat{H}_{I}|i> \right|^{2}(2\pi)\delta(\epsilon_f-\epsilon_i+\omega_{p^\prime p})\,,
		\end{aligned}
	\end{equation}
	where $p_{i}$ is the thermal distribution of the initial state $|i>$. Using the fluctuation-dissipation theorem, the correlation function $S^{\hat{H}^{\dagger}_{I}\hat{H}_{I}}(\omega)$ can be expressed with the zero-temperature approximation as 
	\begin{equation}
	\begin{aligned}		S^{\hat{H}^{\dagger}_{I}\hat{H}_{I}}(\omega) =2V\left|V_{\rm{cou}}(\boldsymbol{Q})\right| \rm{Im}\left[\frac{-1}{\epsilon(\boldsymbol{Q},\omega)}\right]\\
	\end{aligned}
\end{equation}
	with $\epsilon(\boldsymbol{Q},\omega)$ being the dielectric function and $V$ being the material volume. It should be noticed that the structure of the electron part in Eq.(\ref{L_EnES_eff}) is the same with that in the Coulomb interaction and the neutrino part will be calculated with the averaging and spin summing as in the relativistic theory due to the relativistic property of neutrinos. Therefore, to calculate the cross section of E$\nu$ES with this technique, we replace the Coulomb potential with the effective potential of E$\nu$ES in Eq.(\ref{L_EnES_eff}) and the correlation function $S^{\hat{H}^{\dagger}_{I}\hat{H}_{I}}(\omega)$ can be expressed as
	\begin{equation}
		\begin{aligned}
			S^{\hat{H}^{\dagger}_{I}\hat{H}_{I}}(\omega) =2V\frac{\left|V_{\rm{E}\nu \rm{ES}}\right|^2}{\left|V_{\rm{cou}}(\boldsymbol{Q})\right|} \rm{Im}\left[\frac{-1}{\epsilon(\boldsymbol{Q},\omega)}\right]\\
		\end{aligned}
	\end{equation}
	According to the above discussion, one can obtain the cross section for E$\nu$ES in the HEG by inserting the correlation function as
	\begin{equation}
		\label{CS_EnES_HEG}
		\begin{aligned}
			\sigma	=-\frac{\Omega G_F^2g_{\alpha;\rm{V}}^2}{\pi \alpha}\int\frac{\dif^3 Q}{(2\pi)^3} \bm{Q}^2  \rm{Im}\left[\frac{-1}{\epsilon(\bm{Q},\omega)}\right] \delta(\omega +\omega_{p^\prime p}) \dif \omega\,.
		\end{aligned}
	\end{equation}
	\begin{figure*}
	\centering
	\includegraphics[width=\textwidth, angle=0]{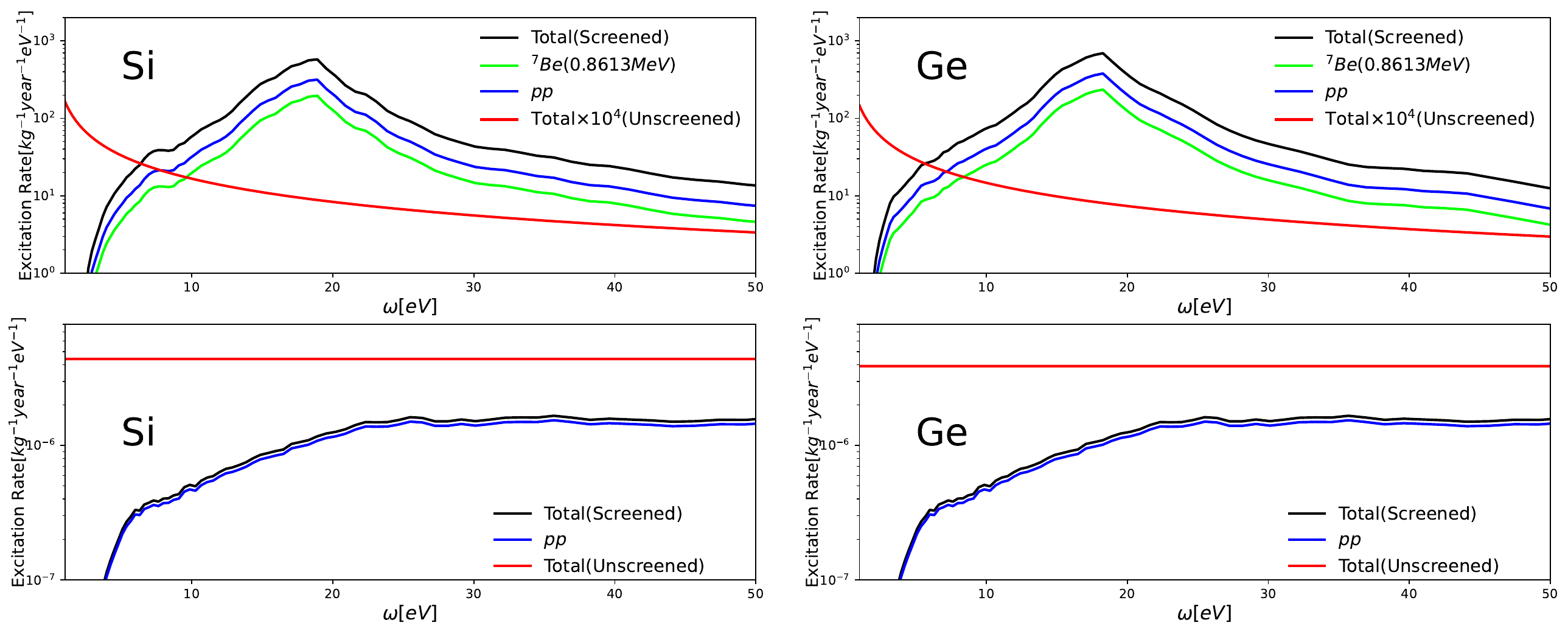}
	\caption{The differential excitation rate in silicon (left) and germanium (right) target induced by solar neutrino with the neutrino magnetic moment $\mu_{\nu}=10^{-11}\mu_{\rm{B}}$ (Top) and in the SM (Bottom). We show the total excitation rate with screening effect with black line and results from independent particle scattering in red line. The main excitation rate comes from the $pp$ neutrino (blue lines). We also show the excitation rate from the $^7\rm{B}$ neutrino with the peak at 0.8613 MeV in the presence of the neutrino magnetic moment, which induces the second largest contribution in this case.}
	\label{fig:spectra}	
	\end{figure*} 
 
	In a crystalline solid, the correlation function is expressed in the reciprocal space periodically and connected to the microscopic dielectric matrix due to the translational symmetry of the crystal lattice. As a result, the transferred momentum $\bm{Q}$ can be split into a reduced momentum $\bm{q}$ in 1 Brillouin Zone (BZ) and a reciprocal momentum $\bm{G}$.  Therefore, with above discussion, the cross section of E$\nu$ES for the HEG in Eq.(\ref{CS_EnES_HEG}) can be extended to the case in the crystalline solid as
	\begin{equation}
	\begin{aligned}
		\sigma	=\frac{\Omega G_F^2g_{\alpha;\rm{V}}^2}{\pi \alpha}\sum_{G}\int_{1\rm{BZ}}\frac{\dif^3 q}{(2\pi)^3} \left|\bm{G}+\bm{q} \right| ^2 \\ \times\rm{Im}\left[\frac{1}{\epsilon_{\bm{G},\bm{G}}(\bm{q},\omega)}\right] \delta(\omega +\omega_{p^\prime p}) \dif \omega\,,
	\end{aligned}
	\end{equation}
	where $\Omega$ is the volume of a cell of the target crystal. In above calculation, we take LFEs into consideration and express the microscopic dielectric matrix element Im$[-\epsilon_{\bm{G},\bm{G}^\prime}^{-1}(\boldsymbol{q},\omega)]$ approximately as its diagonal components Im$[-\epsilon_{\bm{G},\bm{G}}^{-1}(\boldsymbol{q},\omega)]$, which is averaged over $\bm{q}$ and $\bm{G}$. For a target with $N_{\rm{cell}}$ crystal cells exposed to a neutrino flux $\Phi(E_{\nu})$, the excitation rate can be written as
	\begin{equation}
		\label{ER_EnEs_R}
		\begin{aligned}
			{\frac{\dif R}{\dif \omega}}=&\frac{N_{\rm{cell}}\Omega G_F^2g_{\alpha;\rm{V}}^2}{\pi \alpha}\int\Phi(E_{\nu})\dif E_{\nu}\sum_{G}\int_{1\rm{BZ}}\frac{\dif^3 q \dif \omega}{(2\pi)^3} \\ 
			&\times\left|\bm{G}+\bm{q} \right| ^2\rm{Im}\left[\frac{1}{\epsilon_{\bm{G},\bm{G}}(\bm{q},\omega)}\right] \delta(\omega +\omega_{p^\prime p}) \dif \cos\theta_{\boldsymbol{p},\boldsymbol{Q}}\\
		\end{aligned}
	\end{equation}
	In this work we assume an isotropic crystal target and it is a straight forward exercise to integrate out the $\delta$ function in terms of Heaviside function $\Theta$ as
	\begin{equation}
		\begin{aligned}
			\int_{-1}^{1}&\delta(\omega +\omega_{p^\prime p}) \dif \cos\theta_{\boldsymbol{p},\boldsymbol{Q}}\\
			\simeq&\frac{1}{\left|\bm{G}+\bm{q} \right|}[\Theta(\left|\bm{G}+\bm{q} \right|-\omega)\Theta(E_{\nu}-\left|\bm{G}+\bm{q} \right|)\\
			&+\Theta(2E_{\nu}-\left|\bm{G}+\bm{q} \right|-\omega)\Theta(\left|\bm{G}+\bm{q} \right|-E_{\nu})]\, .
		\end{aligned}
	\end{equation}
	\begin{figure*}
		\centering
		\includegraphics[width=\textwidth, angle=0]{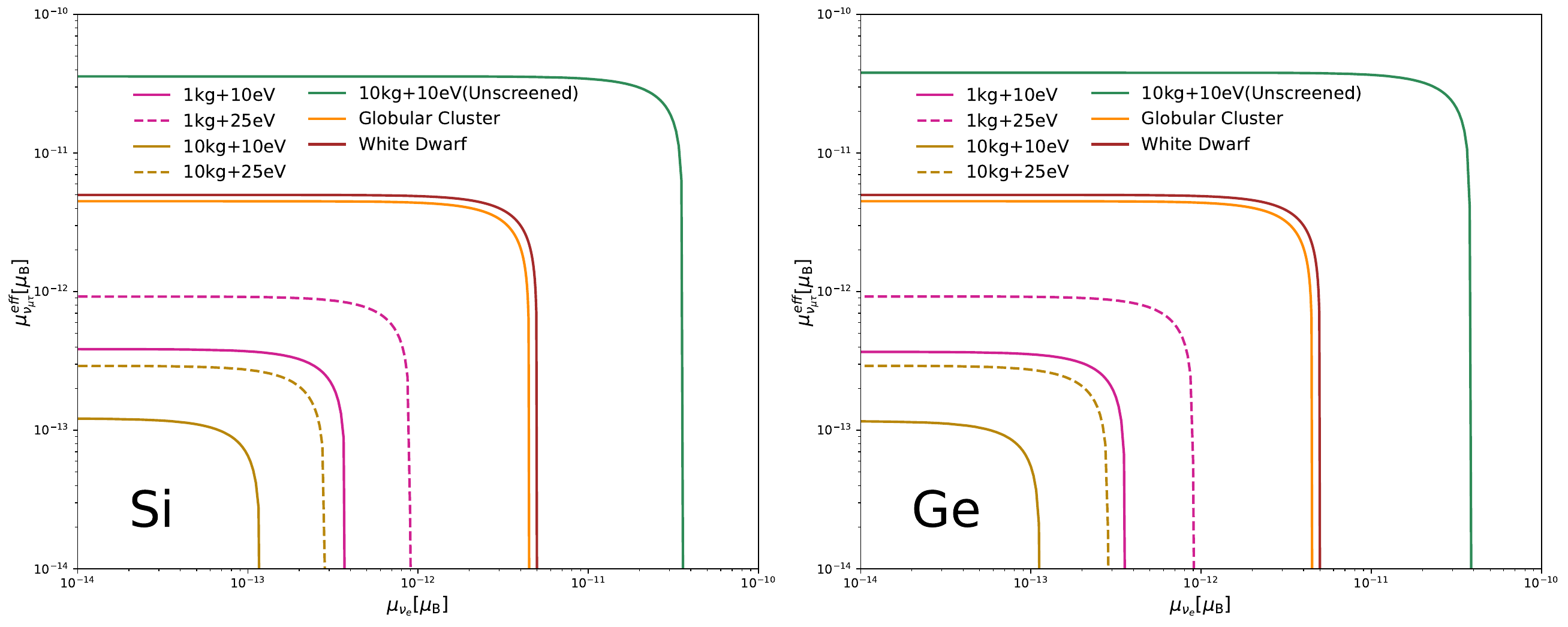}
		\caption{Constraints for the neutrino magnetic moments with screening effect under 50 eV based on silicon (left) and germanium (right) experiment sets of different target mass and threshold. We also show the constraints without screening effect based on 10 kg target and 10 eV threshold. We employ a flat background of 100 $\rm{keV}^{-1}\rm{t}^{-1}\rm{yr}^{-1}$ and an ideal efficiency of 100\% and set the experiment time to be one year. We also show the constraints from the astrophysical observation of the globular cluster M5~\cite{PhysRevLett.111.231301} and the white dwarf~\cite{C_rsico_2014}.}
		\label{fig:limit}	
	\end{figure*}
	With the above relation, the excitation rate in Eq.(\ref{ER_EnEs_R}) can be written in terms of the transferred momentum $\left|\bm{G}+\bm{q} \right|$ and the dielectric matrix element $\epsilon_{\bm{G},\bm{G}}(\bm{q},\omega)$ with a more clear structure
	\begin{equation}
		\label{ER_EnEs_P}
		\begin{aligned}
				{\frac{\dif R}{\dif \omega}}=&\frac{N_{\rm{cell}}\Omega G_F^2g_{\alpha;\rm{V}}^2}{\pi \alpha}\int\Phi(E_{\nu})\dif E_{\nu} \\ 
				&\times\sum_{G}\int_{1\rm{BZ}}\frac{\dif^3 q \dif \omega}{(2\pi)^3}\left|\bm{G}+\bm{q} \right| \rm{Im}\left[\frac{1}{\epsilon_{\bm{G},\bm{G}}(\bm{q},\omega)}\right] \\
				&[\Theta(\left|\bm{G}+\bm{q} \right|-\omega)\Theta(E_{\nu}-\left|\bm{G}+\bm{q} \right|)\\
				&+\Theta(2E_{\nu}-\left|\bm{G}+\bm{q} \right|-\omega)\Theta(\left|\bm{G}+\bm{q} \right|-E_{\nu})]\,.
		\end{aligned}
	\end{equation}
	It should be noticed that the above expression includes the momentum transfer term $\left|\bm{G}+\bm{q} \right|$, which will enhance the response of the target crystal at high values of the momentum transfer.

	For the contribution of the neutrino magnetic moment, we can obtain the excitation rate in a similar way after replacing the effective potential with $V_{\nu,\rm{mag}}$ in Eq.(\ref{L_mag_eff})
	\begin{equation}
		\begin{aligned}
			{\frac{\dif R}{\dif \omega}}=& 32N_{\rm{cell}}\Omega \mu^{2}_{\nu}\frac{\pi\alpha }{m^2_e} \int\Phi(E_{\nu})E^2_{\nu}\dif E_{\nu}\\
				&\times\sum_{G}\int_{1\rm{BZ}}\frac{\dif^3 q \dif \omega}{(2\pi)^3}\frac{1}{\left|\bm{G}+\bm{q} \right|^3} \rm{Im}\left[\frac{1}{\epsilon_{\bm{G},\bm{G}}(\bm{q},\omega)}\right] \\
			&[\Theta(\left|\bm{G}+\bm{q} \right|-\omega)\Theta(E_{\nu}-\left|\bm{G}+\bm{q} \right|)\\
			&+\Theta(2E_{\nu}-\left|\bm{G}+\bm{q} \right|-\omega)\Theta(\left|\bm{G}+\bm{q} \right|-E_{\nu})]\\
		\end{aligned}
	\end{equation}
	In this case, the excitation rate includes the momentum transfer term $1/\left|\bm{G}+\bm{q} \right|^3$ that will significantly enhance the response of target at lower values of the  momentum transfer, which would help us to improve the limit on the neutrino magnetic moment.
 
	\paragraph{\textbf{Constraints on the Neutrino Magnetic Moment.}}
In this work, we choose germanium and silicon, which are the most popular materials for the semiconductor detectors of DM direct detection, as the target materials of this work {and} employ the {corresponding} dielectric function {calculated with the time-dependent density function theory} in Refs.~\cite{Knapen:2021bwg,Knapen:2021run}.

In Fig.~\ref{fig:spectra}, we show the differential excitation rates of the E$\nu$ES induced by solar neutrinos with (upper panels) and without (lower panels) the neutrino magnetic moment. The left and right panels are for the materials of germanium and silicon respectively. {Here solar neutrinos are used for illustration since they are the most intensive natural neutrino source on the Earth.}
In the upper panels, we show the differential excitation rates with $\mu_{\nu}=10^{-11}\mu_{\rm{B}}$, which is similar to the current limits from the LUX-ZEPLIN~\cite{AtzoriCorona:2022jeb} and XENON~\cite{XENONCollaboration:2022kmb,XENON:2020rca} experiments. {The unscreened results as shown in red lines are dominated by the term that inversely proportions to the recoil energy.} {With} the dominant contribution from $pp$ and $^7\rm{B}$ neutrinos, {the screened results of the black lines show a significant peak induced by the plasmon from the material {response}}. As mentioned above, the term $1/\left|\bm{G}+\bm{q} \right|^3$ from the contribution of the neutrino magnetic moment dramatically enhances the response of the electron at lower values of the momentum transfer, where the plasmon peak is much more significant since the plasmon is long-lived in this range~\cite{PhysRevD.101.123012}.
	
In the bottom panels, we show the differential excitation rates in the SM. The {screened results}, which are significantly lower, heavily rely on the electron response and is dominated by $pp$ neutrinos even more than that with the neutrino magnetic moment. Below 5 eV there is nearly no {observable} response and so is the excitation rate. There is no peak in this case since the $\left|\bm{G}+\bm{q} \right|$ term in Eq.~(\ref{ER_EnEs_P}) enhances the response at high momentum transfer, where the plasmon peak is also not significant because the plasmon has a large decay width in this range for its dispersion matching into kinematically-accessible single electron-hole excitations~\cite{PhysRevD.101.123012}.
	
In Fig.~\ref{fig:limit} we illustrate the constraints on the neutrino magnetic moment including the screening effect {for different detector setups}, with $\mu_{\nu_e}$ being the electron neutrino magnetic moment and $\mu_{\nu_{\mu\tau}}^{\rm eff}\simeq\sqrt{ 0.49 \mu_{\nu_{\mu}}^2+0.51 \mu_{\nu_{\tau}}^2}$ being the effective parameter for $\mu$ and $\tau$ neutrinos~\cite{Yue:2021vjg}. Since the background study at such low energy is lacking, we employ a flat background of 100 $\rm{keV}^{-1}\rm{t}^{-1}\rm{yr}^{-1}$ based on the discussion of SuperCDMS in Ref.~\cite{SuperCDMS:2016wui}. Nowadays semiconductor detectors have already achieved a high efficiency for ionization signals and it is a convenient approximation to employ an ideal efficiency of 100\% for illustration. 
From the figure, one can find that {a 10 eV} threshold {capable of detecting} the plasmon {effects} would dramatically improve the sensitivity. We also note that increasing the exposure from 
1 kg$\cdot$yr to 10 kg$\cdot$yr will also improve the sensitivity by a factor of 3. Finally one would achieve an unprecedented sensitivity of $1\times10^{-13}\mu_{\rm{B}}$ using semiconductor detectors with the screening effect, which is much better than 
the constraints from the astrophysical observation of the globular cluster M5~\cite{PhysRevLett.111.231301} and the white dwarf~\cite{C_rsico_2014}.
Note that the sensitivity without the screening effect is at the level of $4\times10^{-11}\mu_{\rm{B}}$ even with 10 kg$\cdot$yr and the 10 eV energy threshold.

	\paragraph{\textbf{Conclusion.}}

In this work, we have developed for the first time the theoretical description of the neutrino electron excitation at low energies in semiconductors including the screening effect based on the fluctuation dissipation theorem, both within the SM and in the presence of the neutrino magnetic moment. 
We have shown that the excitation behaviors of the E$\nu$ES process in semiconductor detectors are  significantly altered because of the screening effect,
and the excitation rates from the neutrino magnetic moment can be dramatically enhanced.
The sensitivity on the neutrino magnetic moment can be significantly improved to the level of $10^{-13}\,\mu_{\rm B}$, which is much better than the current limits from laboratory and astrophysical probes, and would be important for the search for new physics beyond the SM.
  
\paragraph{\textbf{Acknowledgements.}}
The authors are very grateful to Zhengliang Liang for helpful discussions on the screening effect and the relative models. This work was supported in part by National Natural Science Foundation of China under Grant Nos.~12075255, 12075254 and 11835013, by the Key Research Program of the Chinese Academy of Sciences under Grant No.~XDPB15.

{
\section{APPENDIX A: APPLICATION OF NON-RELATIVISTIC EFFECT FIELD THEORY}
	In this appendix, we provide a detailed application of the NR EFT framework from Ref.~\cite{Mitridate:2021ctr} on the E$\nu$ES process in the SM. The case in the presence of the neutrino magnetic moment can be calculated in a similar way. 
 
 To begin with, apart from the coupling to incident neutrinos, the electrons in semiconductors also couple to the background electromagnetic (EM) field, which can be described with the SM Lagrangian
\begin{equation}
	\mathcal{L}_{{\rm EM}}=\bar{\phi}_{e}[i\gamma^{\mu}(\partial_{\mu}+ieA_{\mu})-m_{e}]\phi_{e}\,.
\end{equation}
In the NR EFT framework, the electron fields can be written in the effective operators 
\begin{equation}
		\label{phiplus}
	\phi_{\pm}(\bm{x},t)=e^{-im_{e}t}P_{\pm}\phi_{\rm NR}(\bm{x},t)\,,
\end{equation}
where $P_{\pm}\equiv(1\pm \gamma^{0})/2$ is the projection operators and $\phi_{\rm NR}$ is the electron field in the non-relativistic theory. With the effective operator $\phi_{\pm}$, the SM Lagrangian can be written as
\begin{equation}
	\begin{aligned}
		\mathcal{L}_{\rm EM}&=\phi_{+}^{\dagger}(i\partial_{t}-eA_{0})\phi_{+}+\phi_{-}^{\dagger}(i\partial_{t}-eA_{0}+2m_{e})\phi_{-}\\
		&+\phi_{+}^{\dagger}i\bm{\gamma}\cdot(\bm{\nabla}-ie\bm{A})\phi_{-}-\phi_{-}^{\dagger}i\bm{\gamma}\cdot(\bm{\nabla}-ie\bm{A})\phi_{+}\,.
	\end{aligned}
	\label{Lagrangian_EM}
\end{equation} 
In this case, the photon field consists of a electrostatic background $A_{\rm bg}$ and a quantum fluctuation $\mathcal{A}^{\mu}$
\begin{equation}
	\begin{aligned}
		A_{0}(\bm{x},t)&=A_{\rm bg}(\bm{x})+\mathcal{A}_{0}(\bm{x},t)\,,\\
		\bm{A}(\bm{x},t)&=\bm{\mathcal{A}}(\bm{x},t)\,,
	\end{aligned}
\end{equation}
After separating the electrostatic background and quantum fluctuation components, we can integrate out the heavy field $\phi_{-}$ with the equation of motion and expand terms including $\nabla^2$ with the non-relativistic Schrödinger equation at the leading order, which is already illustrated in detail in Ref.~\cite{Mitridate:2021ctr}. Then we have
\begin{equation}
	\begin{aligned}
		\mathcal{L}^{{\rm eff}}_{{\rm EM},\mathcal{A}}=&-e\mathcal{A}\phi_{+}^{\dagger}\phi_{+}-\frac{ie}{2m_e}\bm{\mathcal{A}}(\phi_{+}^{\dagger}\overleftrightarrow{\bm{\nabla}}\phi_{+})\\
		+\frac{e}{2m_e}&(\bm{\nabla}\times\bm{\mathcal{A}})\cdot(\phi_{+}^{\dagger}\bm{\Sigma}\phi_{+})-\frac{e^2}{2m_e}\bm{\mathcal{A}}^2\phi_{+}^{\dagger}\phi_{+}\,.
	\end{aligned}
\end{equation}
The Lagrangian for E$\nu$ES at low energies in Eq.~(\ref{Lagragian_EvES_VA}) includes both vector and axial-vector terms and we need to match onto the NR EFT framework in different ways due to their different Lorentz structures. To include the vector contribution of the E$\nu$ES Lagrangian with a background electromagnetic field, we make the following replacement
\begin{equation}		
e\mathcal{A}^{\mu}\rightarrow e\mathcal{A}^{\mu}-\frac{G_{\rm F}}{\sqrt{2}}g_{\alpha;\rm{V}}L^{0}\,.
\end{equation}
Then after a similar calculation we obtain the effective Lagrangian for E$\nu$ES with the contribution of the background electromagnetic field subtracted
\begin{equation}
	\begin{aligned}
		\mathcal{L}^{\rm eff}_{\rm{E}\nu \rm{ES},\rm{V}}=-i\sqrt{2}G_{\rm F} g_{\alpha;\rm{V}}\bar{\nu}_{{\alpha},L}\gamma^{0}\nu_{{\alpha},L}\phi_{+}^{\dagger}\phi_{+}\,.
	\end{aligned}
\end{equation}
For the axial vector part, since there is no such structure in the electromagnetic part, we need to include such structure into the equation of motion. However, it is reasonable to substitute the equation of motion in this case with that of Eq.(\ref{Lagrangian_EM}) and integrate $\phi_{-}$ out in a similar way as the vector case since we only consider the terms at the leading order~\cite{Mitridate:2021ctr}.
Therefore we can obtain the leading order Lagrangian as
\begin{equation}
	\begin{aligned}
		\mathcal{L}^{\rm eff}_{\rm{E}\nu \rm{ES},\rm{A}}=-i\frac{G_F}{\sqrt{2}}g_{\alpha;\rm{V}}\bm{L}\phi_{+}^{\dagger}\bm{\Sigma}\phi_{+}\,.\\
	\end{aligned}
\end{equation}
Since the spatial contribution $\bm{L}$ can be approximately neglected compared to the temporal contribution $L^0$, the terms with the spatial contribution will not be included in the leading order Lagrangian. The axial part can also be neglected compared to the vector part at the leading order. 

As a result, the effective Lagrangian of E$\nu$ES can be written with only the vector contribution as
	\begin{equation}
	\mathcal{L}_{\rm{E}\nu \rm{ES}}=-i\sqrt{2}G_Fg_{\alpha;\rm{V}}\bar{\nu}_{{\alpha},L}\gamma^{0}\nu_{{\alpha},L}\phi_{+}^{\dagger}\phi_{+}\,.
	\end{equation}
}

\bibliography{main}

\end{document}